\documentclass[prb,twocolumn,superscriptaddress,amsmath,amssymb]{revtex4-1}
\usepackage{graphicx,times}
\usepackage{epstopdf}
\usepackage{wasysym}
\usepackage{color}

\newcommand{\ket}[1]{\left|{#1}\right>}
\newcommand{\bra}[1]{\left<{#1}\right|}
\newcommand{\tr}[1]{\textnormal{tr}{\left\{#1\right\}}}

\newcommand{\ML}{\widehat{\rho}_\textsc{ml}}
\newcommand{\pML}[1]{\widehat{p}^{\,(\textsc{ml})}_{#1}}
\newcommand{\SML}[1]{S^{(\textsc{ml})}_{#1}}

\newcommand{\CVerr}{\text{PrErr }}
\newcommand{\TRUE}{\rho_\text{true}}
\newcommand{\TARG}{\rho_\text{targ}}

\newcounter{err}
\newcommand{\next}{\refstepcounter{err}\theerr}

\begin{document}

\title{Crystallizing highly-likely subspaces that contain an unknown quantum state of light}

\author{Yong Siah Teo}
\affiliation{Department of Optics, Palack{\'y} University, 17. listopadu 12, 77146 Olomouc, Czech Republic}
\email[Correspondence and requests for materials should be addressed
to  Y.~S.~T. e-mail: ]{yong.teo@upol.cz}
\author{Dmitri Mogilevtsev}
\affiliation{Institute of Physics, Belarus National Academy of Sciences, F.~Skarina Ave. 68, Minsk 220072 Belarus}
\affiliation{Centro de Ci{\^e}ncias Naturais e Humanas, Universidade Federal do ABC, Santo Andr{\'e},  SP, 09210-170 Brazil}
\author{Alexander Mikhalychev}
\affiliation{Institute of Physics, Belarus National Academy of Sciences, F.~Skarina Ave. 68, Minsk 220072 Belarus}
\author{Jaroslav {\v R}eh{\'a}{\v c}ek}
\affiliation{Department of Optics, Palack{\'y} University, 17. listopadu 12, 77146 Olomouc, Czech Republic}
\author{Zden{\v e}k Hradil}
\affiliation{Department of Optics, Palack{\'y} University, 17. listopadu 12, 77146 Olomouc, Czech Republic}

\begin{abstract}
In continuous-variable tomography, with finite data and limited computation resources, reconstruction of a quantum state of light is performed on a finite-dimensional subspace. No systematic method was ever developed to assign such a reconstruction subspace---only \emph{ad hoc} methods that rely on hard-to-certify assumptions about the source and strategies. We provide a straightforward and numerically feasible procedure to uniquely determine the appropriate reconstruction subspace for any given unknown quantum state of light and measurement scheme. This procedure makes use of the celebrated statistical principle of maximum likelihood, along with other validation tools, to grow an appropriate seed subspace into the optimal reconstruction subspace, much like the nucleation of a seed into a crystal. Apart from using the available measurement data, no other spurious assumptions about the source or \emph{ad hoc} strategies are invoked. As a result, there will no longer be reconstruction artifacts present in 
state reconstruction, which is a usual consequence of a bad choice of reconstruction subspace. The procedure can be understood as the maximum-likelihood reconstruction for quantum subspaces, which is an analog to, and fully compatible with that for quantum states.
\end{abstract}

\date{\today}

\begin{widetext}
\maketitle
\end{widetext}

One of the scientifically established tenets in quantum mechanics is the ability to reconstruct any quantum state of an arbitrary quantum source \cite{ysteo_book,rehacek_book}. Maturation of theoretical and experimental techniques in quantum tomography for continuous-variable (CV) measurements is of top priority for practical certifications in optical quantum cryptography \cite{cvQC1,cvQC2,cvQC3,cvQC4,cvQC5}, optomechanics \cite{optomech1,optomech2}, quantum metrology \cite{qmetro1,qmetro2} and other quantum computation protocols \cite{cvtelep1,cvtelep2,cvdc1,cvdc2,cvclone1,cvclone2}.

Since measurement data and computation resources are always finite, the reconstruction of any quantum state of light, which in principle resides in an infinite-dimensional Hilbert space, is always performed on a finite-dimensional subspace. An unsolved problem in CV tomography is an objective systematic search for the appropriate reconstruction subspace. Ideally, an observer would hope for an analytical reasoning that leads to the optimal reconstruction subspace that minimizes some sort of tomographic accuracy measure. This thinking is, in some sense, naive as such an optimal subspace would always depend on the measurement scheme and the true quantum state of the source, an element that is certainly unknown to the observer. Furthermore, the positivity constraint on quantum states forbids any straightforward analysis on the problem.

Despite the aforementioned difficulties, there exist numerous studies on alternative solutions to this problem. These studies, nonetheless, involve making some assumptions about the source. If the observer knows, usually with low to moderate levels of confidence, that the source emits no more than $D_\text{rec}$ photons, then in principle, she can prepare a set of CV measurement outcomes that is \emph{informationally complete} on the $D_\text{rec}$-dimensional Hilbert subspace \cite{sych_HOMO}. The {\it maximum-likelihood} (\emph{ML}) method \cite{ml}, for instance, can be used to reconstruct the state on this subspace based on the measurement data. However, in \cite{mlme} it was shown that such a simple approach often gives estimators that are far away from the true state $\TRUE$, especially when there are features in high-dimensional sectors that are not obvious from simple deductions with mean photon numbers. In the same articles, the technique of \emph{maximum-likelihood-maximum-entropy} (MLME) was used 
to reconstruct states on subspaces larger than the tomographic coverage of the measurement outcomes to reveal genuine quantum-state features and reduce reconstruction artifacts on average.

Recently, methods employed in classical statistical-model selection were
used to localize the signal (see, for example, Refs.~\cite{usami,guta2012,yin}). These methods involve the consideration of
the popular Akaike criterion and the Bayesian
information criterion to penalize the likelihood function for the problem and
restrict models for up to a certain number of parameters. However, applications of these methods on
quantum-state reconstruction is not straightforward because of the quantum positivity constraint \cite{anraku,king}.  Moreover, there is no guarantee that the conditions for the validity of these commonly-used criteria are met, since a truncation of the subspace introduces additional systematic errors. Other methods of choosing reconstruction subspaces include the utilization of other prior knowledge about the source and assigning a partial dependence of the subspace dimension $D_\text{rec}$ on the number of measurement settings or groups of outcomes \cite{artiles}.

In what follows, we shall present a systematic and practical procedure to locate subspaces that highly-likely contain a given unknown quantum state of light that is completely free of any \emph{ad hoc} assumptions about the source. In a nutshell, this procedure makes use of the ML strategy to define an initial reconstruction subspace of low-dimension and gradually evolve the seed subspace to a reconstruction subspace of a stipulated dimension $D_\text{rec}$---much like a typical nucleation process in the formation of crystals. The termination of the ML nucleation process, and the subsequent determination of $D_\text{rec}$, is governed by the procedure of cross-validation, which is a prototypical statistical validation tool that ensures the reliability and predictive power of the resulting ML state estimator. This numerical nucleation process, which is naturally compatible with ML state estimation \cite{ysteo_book,rehacek_book}, makes use of \emph{only} the acquired measurement data in an experiment and does 
not depend on any spurious and/or \emph{ad hoc} assumptions about the quantum source. The underlying physical reason is that all encoded information in the data reflects the features of the unknown quantum state, albeit with some statistical fluctuation, and can thus be systematically extracted to obtain the optimal reconstruction subspace and state estimator.

Without loss of generality, we shall assume here that the data associated with the continuous-variable quantum measurement, although finite, are sufficiently large enough such that statistical fluctuation is minimized within typical experimental means. In this situation, the relevant reconstruction error of interest is primarily influenced by the choice of reconstruction subspace.\\[1ex]

\noindent
{\bf Results}\\[1ex]
\noindent
{\bf The ML subspace nucleation process.}~Suppose that the observer chooses to reconstruct the true quantum state $\TRUE$ of the source using the ML method from a set of data. In CV tomography, the data are event occurrences $\sum_jn_j=N$ of $N$ sampling events (say voltage detection) collected with a measurement described by a set of probability operator measurement (POM) $\sum^M_{j=1}\Pi_j=1$ consisting of $M$ outcomes $\Pi_j\geq0$. In this scenario, it is natural to consider an assignment of the reconstruction subspace that is compatible with the ML principle. Clearly, if the desired ML estimator $\ML$ is the one that maximizes the log-likelihood function of $\rho$ for the data,
\begin{equation}
\log\mathcal{L}(\{n_j\};\rho)=\sum^M_{j=1}n_j\log p_j\,,\,\,\text{where}\,\,p_j=\tr{\rho\Pi_j}\,,
\end{equation}
the reconstruction subspace should then be a subspace of a certain dimension $D_\text{rec}$ that optimizes this log-likelihood function. The problem now reduces to deciding the appropriate value of $D_\text{rec}\geq2$ and searching for the optimal subspace of this dimension.

Since all we have are the measurement data $\{n_j\}$, the most straightforward way to carry out the subspace search is \emph{subspace nucleation}. Such a numerical nucleation process involves the surveillance of all possible $d$-dimensional discrete subspaces of some large Hilbert space of dimension $D_\text{lim}$ that defines some limit for the state reconstruction. All the $L=\binom{D_\text{lim}}{d}$ subspaces can be represented by a set of $L$ projectors $\{S_{l,d}\}^L_{l=1}$ that are diagonal in the computational basis and consist of $D_\text{lim}-d$ zeros and $d$ ones. For any given operator $A$, its suboperator $A_{l,d}$ in the $l$th subspace is conveniently expressed as $A_{l,d}=S_{l,d}AS_{l,d}$. Analogous to nucleation in crystal formation, subspace nucleation begins with a seed subspace of a certain smallest pre-chosen dimension $d$. For the purpose of illustration, we take $d=2$. The qubit subspace $\SML{2}$ appropriate for seeding the nucleation process is the one corresponding to the two-
dimensional $\ML$ of the largest maximal log-likelihood out of all possible projectors $S_{l,2}$. The subspace begins to grow along the trajectory of largest likelihood increment. The next optimal subspace to choose would be the subspace $\SML{4}=\SML{2}+S_{2}$, where $S_{2}$ is the optimal orthogonal subspace to $\SML{2}$, such that the corresponding four-dimensional $\ML$ yields the largest maximal log-likelihood. Nucleation continues, this time establishing the next larger optimal subspace $\SML{6}=\SML{4}+S_{2}$ such that, again, $S_{2}\SML{4}=0$ and the corresponding six-dimensional $\ML$ gives the largest maximal log-likelihood, and so on.

In this way, the reconstruction subspace matures in the direction of maximal sequential increase in the log-likelihood. Since the process evaluates the (log-)likelihood and maximizes it over subspaces, this process is entirely equivalent to a ML subspace estimation, which is fully analogous to a ML state estimation. The data alone contain all hidden signatures of the relevant subspace segments and the structures thereon, all of which are revealed by nothing else but the log-likelihood function. In the limit of large $N$ of sampling events, which is an achievable commodity in homodyne tomography, for instance, these signatures accurately reflect those of $\TRUE$. This function thus serves as the only important objective function following which subspace crystallization takes place. No additional spurious and/or \emph{ad hoc} assumptions about the source. We have therefore established a fully objective numerical procedure for assigning reconstruction subspaces that is compliant with ML state estimation.

Computationally, the ML subspace nucleation process is a continuous iteration of the following simple numerical steps over $k$, starting with the smallest optimal $d$-dimensional seed subspace defined by $\SML{d}$ at $k=1$ and proceeding till $k=\kappa$ that defines the final reconstruction-subspace dimension $D_\text{rec}$:\\[0.1cm]
\noindent\hfill%
\begin{tabular}{@{}r@{--\ }p{226pt}}
\next&
In the $k$th step, look for the full set of operators \mbox{$\mathcal{S}_\perp=\{S_{l,d}\}$} that are orthogonal to $\SML{kd}$.
\\[0.8ex]\next&
Set $\SML{(k+1)d}=\SML{kd}+S_d$ with $S_d\in\mathcal{S}_\perp$ that maximizes $\log\mathcal{L}(\{n_j\};\ML)$, where $\ML$ resides in the subspace defined by $\SML{(k+1)d}$.
\end{tabular}\\[0.1cm]
The {\bf Methods} section provides more explicit details on the numerical procedure.\\[1ex]

\noindent
{\bf Criterion for nucleation termination.}~The final task is now to decide on the reasonable value of $D_\text{rec}$. Various statistical tools are available for this purpose, the choice of which depends on the application of the statistical operator $\ML$. Typically, the estimator $\ML$ is used for statistical prediction of probability distributions for future measurement schemes. Some measure of predictive power for the estimator is hence necessary to judge if the related subspace acquired from the nucleation process is sufficiently accurate in data prediction. Physically, the reconstruction subspace that best predicts data should be the largest possible subspace that tomographically covers all the possible datasets (infinite-dimensional in principle). In practice, however, all resources are finite and some sort of statistical certification is necessary to judge if the estimator of finite data that resides in a finite subspace is predictive enough.

\begin{figure}[h!]
\centering
\includegraphics[width=0.99\columnwidth]{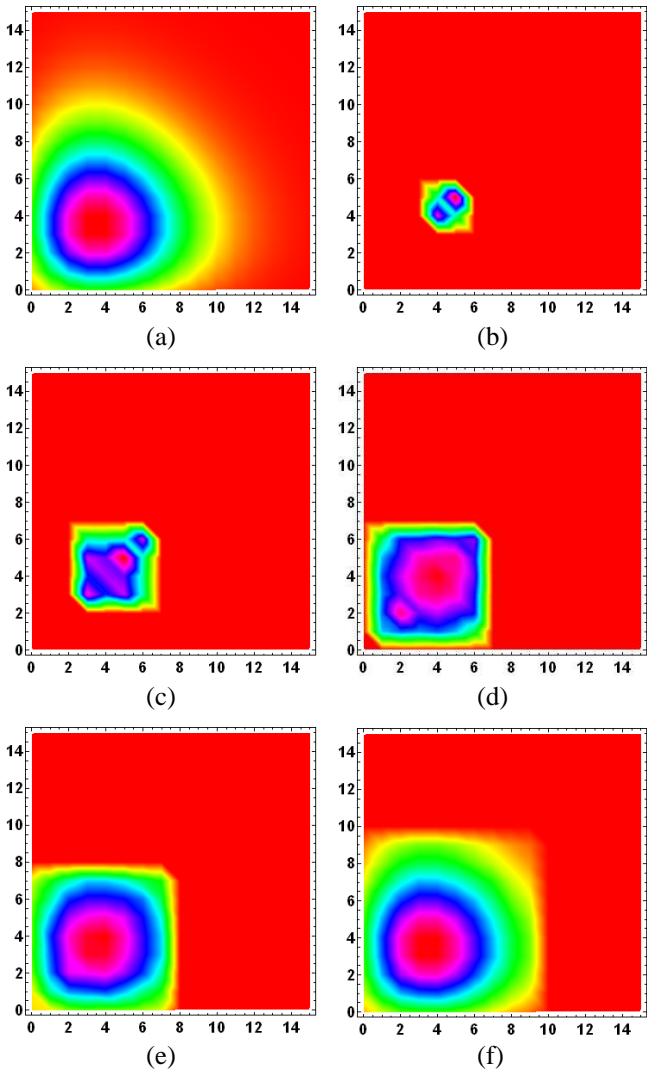}
\caption{\label{fig:MLcrystal_coh}Subspace nucleation process from one set of data for (a) the coherent state defined by $\ket{\alpha}$ of mean-photon number equal to $|\alpha|^2=4$, projected onto the 16-dimensional Hilbert space for visualization. The seed subspace is of dimension $d=2$. Subspaces of (b)~$D_\text{rec}=2$, (c)~4, (d)~6, (e)~8 and (f)~10 that respectively maximize the log-likelihood are shown here for $M=1000$ measurement outcomes. The interpolated hue for each integer coordinate (position of the matrix element) in the plots visually indicates the relative magnitudes of neighboring matrix elements of the real parts of all quantum states in the computational basis. Here the ten-dimensional optimal ML subspace already captures most of the important features of the state.}
\end{figure}

\begin{figure}[h!]
\centering
\includegraphics[width=1\columnwidth]{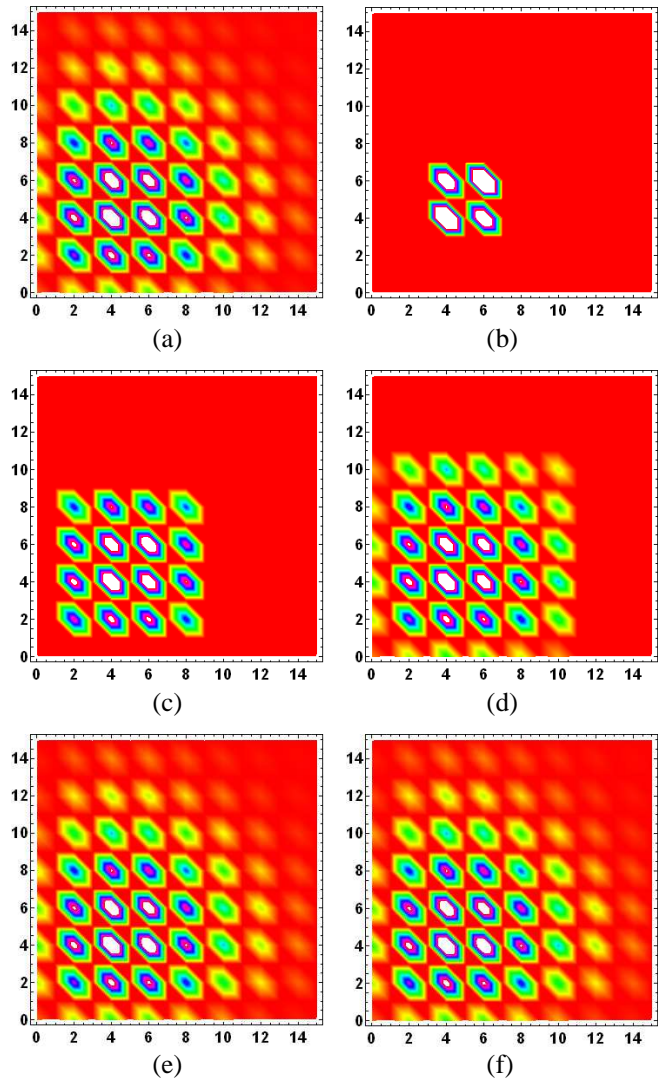}
\caption{\label{fig:MLcrystal_ecoh}Subspace nucleation process from one set of data for (a) the even coherent state defined by $\mathcal{N}(\ket{\alpha}+\ket{-\alpha})$ with $\alpha=\sqrt{5}$ and a proper normalization. All other figure specifications are as described in Fig.~\ref{fig:MLcrystal_coh}. For this state, the eight-dimensional optimal ML subspace is sufficient for a rather accurate ML reconstruction.}
\end{figure}

\begin{figure*}[htp]
\centering
\includegraphics[width=\textwidth]{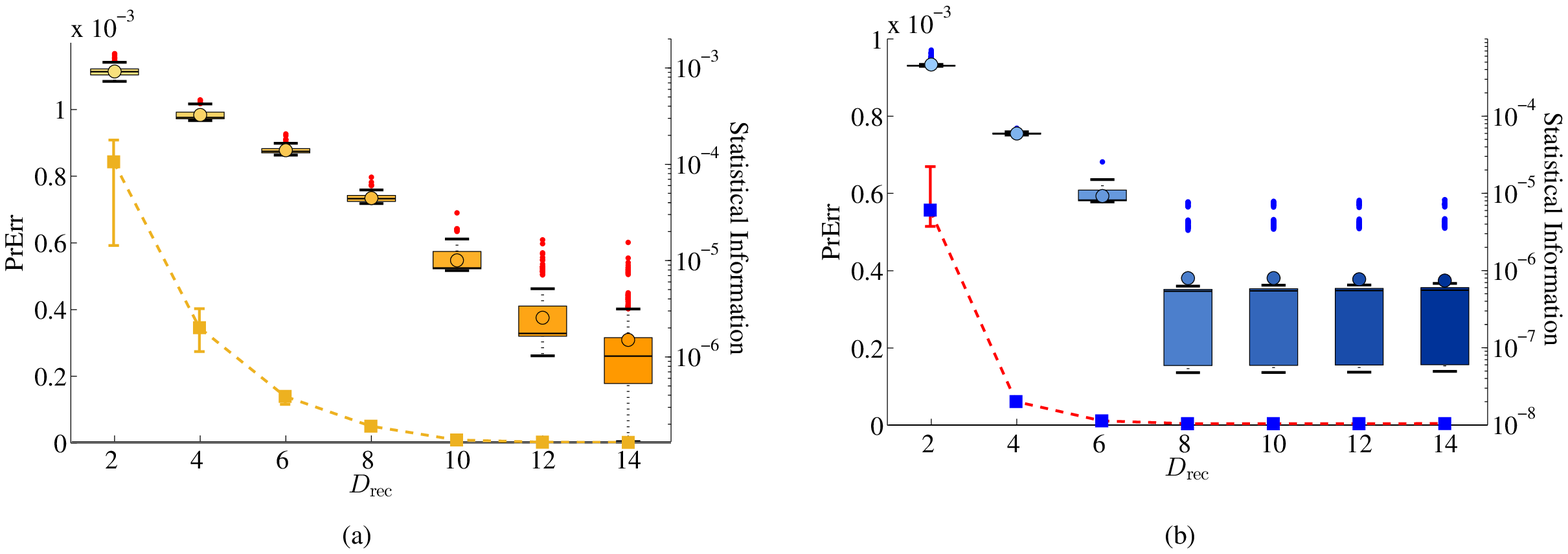}
\caption{\label{fig:StatsInfo}Superimposed plots of the $\CVerr$ values with error bars (squares and dashed lines plotted on a linear scale) and corresponding statistical information about each value (plotted on a log scale) against $D_\text{rec}$, respectively for (a)~the coherent state and (b)~the even coherent state. Each error bar is computed from the relevant bootstrap distribution. Small error bars are not visible in the figure. The statistical information for each value of $D_\text{rec}$ is based on a bootstrap distribution of 500 Monte-Carlo-generated $\CVerr$ values. This information includes the first and third quartiles of these points (respectively the bottom and top edges of the rectangle), the median or second quartile (solid line in the rectangle), the mean (circle), the lowest datum still within 1.5 interquartile range (IQR) of the first quartile and the highest datum still within 1.5 IQR of the third quartile (respectively the bottom and top solid lines of the whisker). Outliers, which are 
outside the whisker, are plotted as vertically-aligned dots.}
\end{figure*}

\emph{Cross-validation} \cite{CrossVal1,CrossVal2,CrossVal3} (and a simplified variant discussed in \cite{CrossVal4}) is a decent certification tool of choice to judge if $\ML$ resides in a large enough reconstruction subspace of reasonable coverage relative to the data of a given POM. Typically, when the estimator $\ML$ fits the data from a POM according to the log-likelihood function, the estimator may not necessarily predict other data from the same POM, or any other POM for that matter, especially in a situation where the reconstruction subspace does not tomographically include the measurement data sufficiently. If, on average, $\ML$ predicts different sets of data of the same POM, then the subspace yields a predictive $\ML$. For demonstrating the principles of cross-validation, we consider a \emph{two-fold} cross-validation strategy and split the $M$ measurement data into two datasets of equal size. Borrowing the language of machine learning, one dataset, the \emph{training set}, is used to obtain $\ML$,
 and the other \emph{testing set} is used to test the predictive power of $\ML$. The roles of both datasets are then switched, and training and testing are performed again. The average chi-square metric between the test data and the ML probabilities,
\begin{equation}
\CVerr=\left.\dfrac{1}{M}\sum^2_{k=1}\sum^{M/2}_{j=1}\dfrac{\left(n_j/N-\pML{j}\right)^2}{\pML{j}}\right|_{k\text{th testing set}}\,,
\end{equation}
describes the predictive power for $\ML$ in terms of the prediction error for the given measurement scheme.

Like all numerical algorithms, there are many ways to terminate the nucleation procedure. The observer may choose to set a pre-chosen tolerance level for PrErr beyond which the procedure stops; or compare the change in the current PrErr value relative to the preceding value and accept the reconstruction if the change falls below certain threshold; or simply repeat the procedure a pre-chosen number of times. The numerical stabilization of PrErr can serve as an indication that continuing the procedure will not give appreciable improvement in the resulting ML~estimator and ML~subspace. Since the value of \CVerr fluctuates for every experimental run, its value for each reconstruction-subspace dimension $D_\text{rec}$ should be accompanied by a statistical quantifier for its reliability. As a typical choice, we shall assign confidence intervals to reflect the level of confidence (or signifcance) for these values. These confidence intervals are calculated using a known method of \emph{bootstrapping} on the $\
CVerr$ values (see {\bf Methods}).\\[1ex]

\noindent
{\bf Numerical Experiments.}~To put the ML subspace nucleation procedure to the test, for a given true state $\TRUE$, we simulate an experimental run for a CV POM involving $M=1000$ random rank-one POM outcomes distributed uniformly according to the Haar measure \cite{karol}. In this run, a total of $N=10^7$ sampling events is measured with the POM and the resulting data are accumulated through Monte Carlo methods. To demonstrate the proposed numerical method, we investigate two such experimental runs, one for a coherent state and another for an even coherent state. Figures~\ref{fig:MLcrystal_coh} and \ref{fig:MLcrystal_ecoh} provide a visualization of the respective nucleation processes for these two states. Figure~\ref{fig:StatsInfo} presents the results of the nucleation process with statistical descriptions for the $\CVerr$ values.

The results obtained with simulated experiments verify the decreasing behavior for the values of the prediction error $\CVerr$ with increasing reconstruction-subspace dimension $D_\text{rec}$. This behavior confirms that, logically, a larger subspace would more adequately accomodate the measurement outcomes and more accurately predict any data derived from these outcomes.\\[1ex]

\noindent
{\bf Practical Aspects Of The Nucleation Methodology.}~{\it Subspace coverage}---In the usual situation, the observer has already an intended target quantum state $\TARG$ for the source in mind before setting up the experiment for a particular quantum protocol. Owing to experimental imperfections, the target state she intends to prepare is never the same as the true state $\TRUE$ to which she asymptotically measures. Nevertheless, if the control of the source is done well, the observer may have reasons to believe that $\TRUE$ should be close to $\TARG$. For a given basis, the reconstruction dimension $D_\text{rec}$, and hence the limit dimension $D_\text{lim}$, should at least be large enough to encompass all the significant matrix elements of $\TRUE$. Usually, the choice of $D_\text{lim}$ is decided from $\TARG$ by trusting that it is close enough to the unknown $\TRUE$.

However, such a gut feeling is not a necessary ingredient to pick the value of $D_\text{lim}$, for the data themselves already contain all encoded information about the quantum-state features. If $D_\text{lim}$ is too small to cover the significant features of $\TRUE$, the data should be able to tell us just that, which they do indeed. One way of capturing the tell-tale signs from the data is to inspect the $\CVerr$ values. If the $\CVerr$ saturates at a value that is large, then this is an indication that the subspace does not cover the state features very well, and the corresponding estimator does not explain the data obtained and will have limited predictive power. In this case, one would need to increase the value of $D_\text{lim}$. The behavior of $\CVerr$ with $D_\text{rec}$ would eventually stabilize for sufficiently large $D_\text{lim}$.

\begin{figure}[t]
\centering
\includegraphics[width=1\columnwidth]{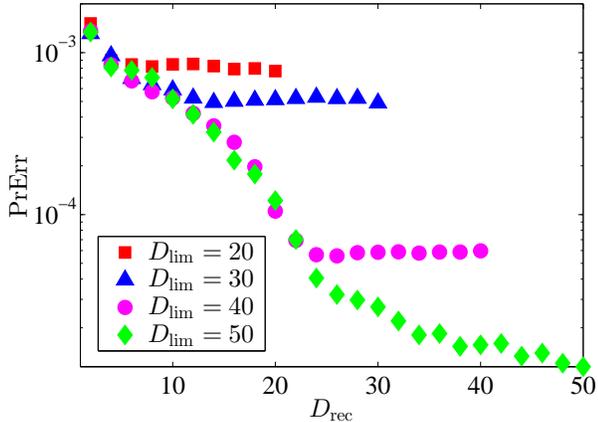}
\caption{\label{fig:cover}A plot of $\CVerr$ (log-scale) against $D_\text{rec}$ for a fixed set of data and various limit dimensions $D_\text{lim}$. The true state is a coherent state of mean photon number 30.}
\end{figure}

As an example, Fig.~\ref{fig:cover} shows the behavior of $\CVerr$ for different values of $D_\text{lim}$, obtained from a fixed set of simulated data of a coherent state with mean photon number 30. As $D_\text{lim}$ increases, the saturation of $\CVerr$ lowers and vanishes for sufficiently large $D_\text{lim}$. In this way, the choice of $D_\text{lim}$ is optimized without the need for a prior belief.\\[1ex]

\noindent
{\it Subspace truncation and rate of convergence}---If the observer insists, she can certainly make use of an educated prior belief for the true state to enhance the ML subspace nucleation procedure in an objective way. To understand how, consider the simple case where $\TRUE$ is the single-photon Fock state $\ket{n=1}\bra{n=1}$. If one carries out the nucleation procedure in the computational basis with (hypothetical) noiseless data, then the procedure will terminate after just one step. This is because already after the very first step, the optimal qubit subspace is, of course, any of the subspaces that covers the $n=1$ sector, and the resulting ML estimator $\ML$ is precisely $\TRUE$ since all other matrix elements are zero in the computational basis.

\begin{figure}[h!]
\centering
\includegraphics[width=0.95\columnwidth]{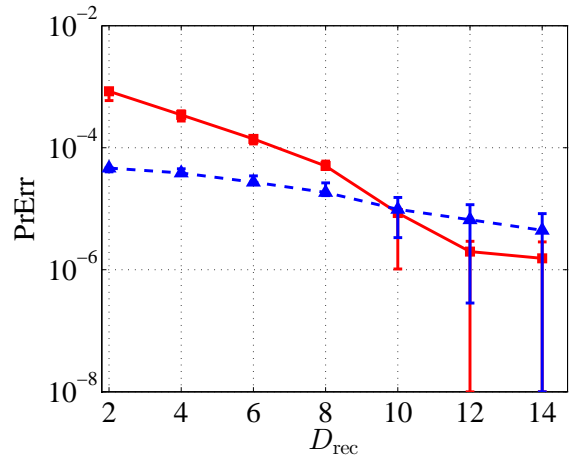}
\caption{\label{fig:basis_opt}A plot of $\CVerr$ (log-scale) against $D_\text{rec}$ for the coherent state discussed in Fig.~\ref{fig:MLcrystal_coh} and a fixed set of simulated data with slight statistical fluctuation. Comparing with the rate of the subspace nucleation process in the standard computational (Fock) basis (red solid lines and square markers), a basis transformation based on the target coherent state of mean photon number five gives a relatively faster convergence (blue dashed line and triangular markers). The intersection of the data-point sequences at $D_\text{rec}=10$ and larger arises from the finite precision of the ML estimation.}
\end{figure}

More generally, for (hypothetical) noiseless data and a rank-one $\TRUE$, if the basis in which subspace truncation of the Hilbert space is performed happens to be the basis with one of the basis kets being the ket of $\TRUE$, then the nucleation procedure yields $\TRUE$ after the very first step, because in this basis all matrix elements are zero except for the one corresponding to the ket of $\TRUE$. This argument is easily extended to any $\TRUE$, in which case truncation in a basis containing all eigenstates of $\TRUE$ will give $\TRUE$ as the estimator in no more than $\text{rank}\{\TRUE\}$ steps. The exact number of steps would depend on the dimension $d$ of the seed subspaces. Even for a realistic situation when the observer has access to only the target state $\TARG$, not $\TRUE$, and real data with statistical fluctuation, finding the right basis with a reasonable $\TARG$ to perform subspace truncation for the nucleation procedure can greatly speed up the nucleation procedure if $\TRUE$ is 
reasonably close to the target state, especially in the limit of large number of sampling events $N$.

It is now clear how the prior belief enters the nucleation procedure---it is simply used to set up the appropriate basis for subspace truncation in order to carry out subspace nucleation with significantly fewer steps. \emph{In no way} is the final state estimator $\ML$ dependent on the prior belief, only the rate of convergence to $\ML$, for the entire nucleation process is still controlled by data inspection alone once the basis is set up. Mathematically, if $U$ is the unitary operator that converts the Fock basis to the appropriate basis, then a basis transformation $\rho\rightarrow U\rho\,U^\dagger$ on quantum states in the optimization routine is entirely equivalent to an inverse basis transformation $\Pi_j\rightarrow U^\dagger\Pi_j U$ on measurement outcomes due to the symmetry in the Born rule. If the observer believes that the target state $\TARG=\ket{\,\,\,}\bra{\,\,\,}$ is the likely candidate for describing the source, she may take this and generate a $D_\text{lim}$-dimensional eigenbasis of $\
TARG$ and construct $U$ out of this eigenbasis. The results in Fig.~\ref{fig:basis_opt} further confirms the possibility of a significant improvement in nucleation convergence to the final optimal subspace and state estimator for a given set of data after a basis transformation.\\[1ex]

\noindent
{\bf Discussions}\\[1ex]
\noindent
We have shown, from these findings, that the maximum-likelihood subspace nucleation procedure is a numerically feasible procedure for obtaining the valid optimal reconstruction subspace that contains the unknown quantum state and, at the same time, maximizes the (log-)likelihood with respect to the measured data. Throughout the procedure, no other assumptions about the source are required. The complete elimination of this requirement turns our proposed procedure into an extremely robust method for real experiments, where such assumptions are sometimes difficult to justify precisely. The reporting of all results on experimental state reconstructions and diagnostics using continuous-variable measurement schemes can now be done more reliably once this restriction is lifted, since the concern of reconstruction artifacts that typically arise from an unsuitable or a suboptimal choice of reconstruction subspace is now out of the picture.

The methods of cross-validation and bootstrapping are used to justify the appropriate size of the optimal reconstruction subspace by investigating its predictive power of future data from the same measurement scheme. Other statistical tools can also be invoked depending on the way the observer uses the resulting quantum-state estimator. In general, all these statistical tools would have to be improved in order to address statistical problems related to the quantum-state space, as the positivity constraint plays an important role in altering the probability distribution of any set of data generated from a quantum state, which would in general be different from its classical counterpart. The study of the implications of the positivity constraint on these statistical methods is beyond the scope of this article.

It should be emphasized that the nucleation methodology is completely general and applicable to quantum-state estimation strategies that are not necessarily invoking the maximum-likelihood principle. Very similar nucleation procedures may be implemented for strategies such as linear-inversion or weighted linear-inversion, for instance. The only difference is that the objective function is no longer the likelihood function, but some other function compatible with the chosen estimation strategy, and the quantum positive constraint can additionally be imposed on all such strategies. The subspace nucleation procedures for these strategies proceed as usual otherwise. The bottom line---the set of data obtained with any CV measurement scheme is the only essential element for an accurate subspace and state reconstruction.\\[1ex]

\noindent
{\bf Methods}\\[1ex]
\noindent
{\bf Detailed numerical procedure for the ML subspace nucleation process.}~In this section, we shall also assume that the largest possible subspace for an efficient reconstruction has dimension defined by some large integer $D_\text{lim}\gg d$---the limit for the state reconstruction. The ``$l$th (reconstruction) subspace of dimension $d$'' can therefore be synonymously understood as the $D_\text{lim}$-dimensional projection operator $S_{l,d}$.

The nucleation process for a particular CV measurement scheme makes use of a list of $L$ seed subspaces $\{S_{l,d}\}^{L}_{l=1}$ of a pre-chosen dimension $d$. As an example, we shall take $d=2$ and $D_\text{lim}=16$, which are the settings for the simulations. Each $(D_\text{lim}=16)$-dimensional projector $S_{l,2}$ is used to compute the maximum log-likelihood with respect to the data. The operators involved in this computation are the state $\rho_{l,d=2}$ and all the POM outcomes $\Pi^{(l,d=2)}_j$ on this particular qubit subspace. More explicitly, for a given $l$, the two-dimensional $\rho_{l,2}$ is simply represented as a two-dimensional positive, unit-trace matrix defined as
\begin{equation}
  \rho_{l,2}=\dfrac{A_{l,2}^\dagger A_{l,2}}{\tr{A_{l,2}^\dagger A_{l,2}}}
\end{equation}
using an auxiliary complex operator $A_{l,2}$. The $j$th outcome $\Pi^{(l,2)}_j$ residing on this subspace is represented by a positive $2\times2$ matrix extracted out of the original $16\times16$ positive matrix $\Pi_j\,\widehat{=}\,M^{(j)}$ describing this outcome. For instance, suppose that $S_{l,2}$ is the $16\times16$ diagonal matrix having only two ``ones'' respectively for the second and fifth diagonal entries. Then the $2\times2$ positive matrix is simply
\begin{equation}
  \Pi^{(l,2)}_j\,\widehat{=}\begin{pmatrix}
                   M^{(j)}_{2,2} & M^{(j)}_{2,5}\\
                   M^{(j)}_{5,2} & M^{(j)}_{5,5}
                  \end{pmatrix}\,.
\end{equation}
Positivity in $\Pi^{(l,2)}_j$ is trivially preserved for every $j$ since this matrix is just the matrix representing $S_{l,d}\Pi_jS_{l,d}$ with only matrix elements on the relevant subspace retained. The sum of all $\Pi^{(l,2)}_j$s is typically not the identity. The ML method regarding such cases are discussed in, for instance, Refs.~\cite{ysteo_book,rehacek_book}. Once the two-dimensional ML estimator $\ML$ for every value of $l$ is computed, the maximal log-likelihood values are then sorted in descending value and the subspace that yields the largest maximal log-likelihood value is then chosen to seed the nucleation process. The set of $\binom{D_\text{lim}=16}{d=2}=120$ projectors is then reduced to the set of $\binom{14}{2}=91$ projectors which now corresponds to a set of subspaces that are orthogonal to the optimal subspace.

The next larger ($d=4$)-dimensional ML subspace is built from this seed by accomodating the optimal qubit seed subspace that is both orthogonal to the current subspace and maximizes the log-likelihood. The subsequent computation is very similar to that described for the previous case, only that $\rho_{l,4}$ and $\Pi^{(l,4)}_j$ are now four-dimensional operators. After this computation, the set of $\binom{14}{2}=91$ projectors is then reduced to the set of $\binom{12}{2}=66$ projectors that are orthogonal to all the selected projectors. The computation rate for this numerical scheme increases with each step as the set of seed subspaces on which ML estimation is performed decreases in size. The procedure continues in this manner until $\ML$ fulfils some fixed criterion that would eventually terminate the nucleation process.\\[1ex]

\noindent
{\bf Cross-validation and bootstrapping.}~\emph{Cross-validation}---If the observer wants to use $\ML$ to predict future measurement data, then the technique of cross-validation is a suitable approach to verify if this ML estimator is predictive. Here, cross-validation is used to verify its predictive power on at least the same measurement scheme. A common technique known as $K$-fold cross-validation involves the splitting of a set of $M$ data into $K$ datasets of equal size. A total of $K-1$ datasets are chosen as training sets to obtain an ML estimator $\ML$. The remaining dataset, the testing set, is then used to test whether $\ML$ gives ML probabilities that are close to these data on average. Other variants of cross-validation exists, some of which possess high computational complexities \cite{CrossVal2}. So far, no systematic studies of cross-validation has been performed for quantum tomography, as such the implications of the positivity constraint, if any, on the quantum-state space are not known. For 
the simulations, $K$ is set to two to ensure that both the training set and testing set are equally large enough. For the specifications of typical homodyne experiments, the binned data are suitable for numerical computation for this value of $K$.

The predictive power of $\ML$ is summarized by the prediction error
\begin{equation}
\CVerr=\left.\dfrac{1}{M}\sum^K_{k=1}\sum^{M/K}_{j=1}\dfrac{\left(n_j/N-\pML{j}\right)^2}{\pML{j}}\right|_{k\text{th testing set}}\,,
\end{equation}
where, without loss of generality, we have assumed that $M$ is divisible by $K$. For a sufficiently large reconstruction subspace, $\CVerr$ would in principle approach zero if not for the slight statistical fluctuations of the measurement data. We mention in passing that in the case where a source drift is present, the true state of the source is no longer stable and describing the source with a single ML estimator using all the measurement data would result in an average bias for $\CVerr$.\\[1ex]
\noindent
\emph{Bootstrapping}---Since $\CVerr$ is statistical, it is in principle necessary to assign some statistical quantifier to it. Any statistical quantifier that describes the reliability of $\CVerr$ would generally require a sample of $\CVerr$ values for each $D_\text{rec}$. With only one set of data, a viable option is to perform bootstrapping on this set of data to generate new sets of pseudodata for the construction of the quantifier. Without the assumption of a model for bootstrapping, the non-parametric bootstrap method is suitable and has been proven to give sample points that follow a distribution close to the population distribution. However, this convergence comes with strings attached, such as the adherence to a list of other assumptions, and these assumptions are not always satisfied for some cases, especially in the presence of the quantum positivity constraint.

A workaround is to suggest that since the $\CVerr$ decreases with increasing $D_\text{rec}$ (if $N$ is large enough that is), we may take the $\ML$ estimator with the smallest $\CVerr$ for bootstrapping---the parametric bootstrapping strategy. This choice of model for the bootstrap data asymptotically guarantees that the resulting bootstrap distribution of random $\CVerr$ values converges to the actual population distribution from the true state as long as $N\gg1$ (typical situation in CV experiments) and $\CVerr\ll1$. The procedure for generating a $\CVerr$ value from a set of pseudodata obtained from a run of parametric bootstrapping is exactly the same as in the case of real data. Parametric bootstrapping is then repeated to accumulate a sample of bootstrap $\CVerr$ values for each $D_\text{rec}$.

The quantifier chosen as an example is the confidence interval that representatively quantifies the confidence level for each $\CVerr$ value. For a given significance level $0<\alpha<1$ that is small, we first compute the $1-\alpha/2$ and $\alpha/2$ percentiles from each bootstrap sample. Upon denoting the percentiles respectively by $\text{PrErr}_{1-\alpha/2}$ and $\text{PrErr}_{\alpha/2}$, the confidence interval is defined as the percentile interval $[2\,\CVerr-\text{PrErr}_{1-\alpha/2},2\,\CVerr-\text{PrErr}_{\alpha/2}]$. The advantage of this interval is that it is computationally efficient and captures approximately some essence of the sample dstributions. A more accurate interval can be acquired by performing a second-level bootstrapping for the standard deviation of each sample, which is often computationally intractable. As an estimate for the spread of $\CVerr$, the percentile confidence interval provides sufficiently reliable information for general purposes. Besides, other statistical information 
is usually needed to supplement this interval for a more thorough data analysis.\\[1ex]

\noindent
{\bf Acknowledgements}\\[1ex]
\noindent
A.~M. and D.~M. acknowledge the support of the National Academy of Sciences of Belarus through the program ``Convergence'', the project 686731 SUPERTWIN of Horizon 2020, and FAPESP grant 2014/21188-0 (D.~M.). Y.~S.~T., J.~{\v R}., and Z.~H. acknowledge the support of the Grant Agency of the Czech Republic (Grant No. 15-03194S), and the IGA Project of the Palack{\'y} University (Grant No. PRF~2015-002).\\[1ex]

\noindent
{\bf Contributions}\\[1ex]
\noindent
D.~M. and Y.~S.~T. contributed to the development of the theory. The manuscript was written by Y.~S.~T., with input and discussions from all other authors. A.~M., J.~{\v R}, and Z.~H. supported and enhanced the research work.\\[1ex]

\noindent
{\bf Additional information}\\[1ex]
\noindent
The author(s) declare no competing interests as defined by Nature Publishing Group, or other interests that might be perceived to influence the results and/or discussion reported in this paper.

\end{document}